\documentclass[granma]{svjour}
\usepackage{graphicx}
\begin{document}
\title{A note on the upward and downward intruder segregation in granular media}
%
\author{Leonardo Trujillo$^{1}$ and Hans J. Herrmann$^{1,2}$}
\institute{
$^1$Physique et M\'ecanique des Milieux H\'et\'erog\`enes\\
\'Ecole Sup\'erieure de Physique et de Chimie Industrielles,\\
10, rue Vauquelin, 75231 Paris Cedex 05, France\\
\\
$^{2}$Institut f\"ur Computeranwendungen 1\\
Universit\"at Stuttgart\\
Pfaffenwaldring 27, 70569 Stuttgart, Germany
}

\date{Received: date / Revised version: date}
%
\maketitle
\begin{abstract}
The intruder segregation dependence on size and density is investigated in the framework of a
hydrodynamic theoretical model for vibrated granular media. We propose a segregation mechanism 
based on the difference of densities between different regions of the granular system, which give origin 
to a buoyant force that acts on the intruder. From the analytic solution of the segregation velocity we can
analyze the transition from the upward to downward intruder's movement.
\end{abstract}
%
The understanding of the behavior of granular segregation is relevant due to their practical 
importance in many industries\cite{Herrmann98,deGennes98}.
When a granular mixture is subject to vertical vibrations under gravity, the grains
tends to segregate with the larger particles at the top of the bed.
Many studies have been devoted to this subject, usually referred as the
``Brazil nut effect"\cite{Rippie64,Ahmad73,Rosato87,Jullien92,Duran93,Knight93,Jullien93,Duran94,Poschel95,Gallas96,Cooke96,Vanel97,Lan97,Caglioti98,Shishodia01,Rosato02}.
There exists a great controversy  concerning the upward to downward segregation in
granular materials\cite{Shinbrot98,Shinbrot00,Hong01,Liffman01,Mobius01,Walliser02}. 
Here we address this problem using a recently
proposed model for intruder size segregation in dry granular media\cite{leo}. This model characterizes
the rise velocity of a large intruder particle immersed in a medium of monodisperse fluidized small
particles. In Ref.\cite{leo} we have proposed a segregation mechanism based on the difference of
densities between different regions of the system, which gives origin to a buoyant force that
acts on the intruder. This force include an {\it Archimedean buoyancy force} due to the differences
between the intruder material density $\rho_I$ and the bed density $\rho_F$, 
${\bf f}_A=(\rho_F - \rho_I)V_I{\bf g}$. Here ${\bf g}$ is the gravity field and 
$V_I = \frac{\Omega_D}{D} r_I^D$ is the $D$--dimensional volume of an intruder with radius
$r_I$. The factor $\Omega_D=2\pi^{D/2}/\Gamma(D/2)$ is the surface area of a $D$--dimensional
unit sphere. 
Also, we include a {\it thermal buoyancy force} caused by density variations of the granular fluid
which comes from differences in the local ``granular temperature'' $\Delta T_g$. The change
in the granular fluid density through the thermal expansion, produced by
the difference of temperatures, is $\rho_F'=\rho_F(1-\alpha \Delta T_g)$, where $\alpha$ is the
thermal expansion coefficient. The thermal contribution to the buoyancy force is 
${\bf f}_T = \Delta \rho_F V_I {\bf g}$, where $\Delta \rho_F = \rho_F' - \rho_F = -\alpha \rho_F \Delta T_g$.
The granular temperature $T_g$ is defined proportional to the mean kinetic energy associated
to the velocity of each particle. The granular temperature difference $\Delta T_g$ is due to the
dissipative nature of the collisions between grains. This difference is due to the fact that the
number of collisions on the intruder surface increases with the size, but the local density
of dissipated energy diminishes. The region with intruder is {\it hotter} than the region
without intruder. This lead  to the thermal buoyancy force that contributes to the intruder's
upward movement. The intruder also experiences a viscous drag force of the granular fluid.
The drag force ${\bf f}_d$ \cite{Zik92,Albert99} is considered to be linear in the velocity of segregation
${\bf u}(t)$, and is like the Stokes' drag force ${\bf f}_d = -6\pi \mu r_I {\bf u}(t)$, where
$\mu$ is the coefficient of viscosity.
Therefore the equation of motion that governs the segregation process is
\begin{eqnarray}
\frac{\Omega_D}{D}r_I^D\rho_I \frac{d {\bf u}(t)}{dt}  & = & \frac{\Omega_D}{D}r_I^D \left[ \rho_I - \rho_F(1+\alpha \Delta T_g) \right] {\bf g}
\nonumber
\\
& & -6\pi \mu r_I {\bf u}(t).
\label{Eqmov}
\end{eqnarray}

We take the reference frame positive in the upward vertical direction. Arranging terms in 
Eq.(\ref{Eqmov}) we find the following differential equation
\begin{equation}
\frac{du(t)}{dt}=\left[ \rho_F (1+ \alpha \Delta T_g)-\rho_I \right]\frac{g}{\rho_I} - \frac{6\pi D \mu}{\Omega_D \rho_I r_I^{D-1}}u(t),
\label{dynamic}
\end{equation}
and the solution is
\begin{equation}
u(t) = t_0 g \left[ \frac{\rho_F}{\rho_I}(1+\alpha \Delta T_g)-1\right]\left(1-e^{-t/t_0} \right),
\label{velocity}
\end{equation}
where the time--scale $t_0$ is
\begin{equation}
t_0\equiv\frac{\Omega_D \rho_I r_I^{D-1}}{6\pi D \mu}.
\label{time}
\end{equation}

The drag force always acts opposite to the intruder velocity. So, the intruder's
upward/downward movement is  exclusively due to the buoyancy forces. For our theoretical
calculations we define the settling velocity $u_s$
\begin{equation}
u_s = t_0 g \left[ \frac{\rho_F}{\rho_I}(1+\alpha \Delta T_g)-1\right].
\label{settling}
\end{equation}
When $u_s>0$ the resulting movement is upward. On the
other hand, if $u_s<0$ the resulting movement is downward.
Our analysis reveals that if $(1+\alpha \Delta T_g)\rho_F/\rho_I>1$, the intruder ascends.
The opposite occurs when $(1+\alpha \Delta T_g)\rho_F/\rho_I<1$. When $u_s=0$ there is no
upward neither downward movement.

The intruder's presence modifies the local temperature of the system due to the collision that 
happen at its surface. In Ref.\cite{leo} we have proposed an analytic procedure to estimate the 
temperature difference among the granular fluid. We can calculate within a sphere of radius $r_0$
the value of the temperature $T_1$ in the granular fluid in  presence of the intruder and
compare it with the temperature $T_2$ in the granular fluid without intruder. 
In both cases we calculate the granular temperatures $(T_1, T_2)$ at a distance $r=r_I$ from the
center of the sphere of radius $r_0$.
See Fig. 1 for
a schematic picture of the regions used to calculate the granular temperature. In the model this 
temperature ratio for two dimensions is given by (see Ref.\cite{leo} for details)

\begin{equation}
\frac{T_1}{T_2} = \left(\frac{I_0(\lambda_Fr_0)[\Theta_{AB}I_0(\lambda_Fr_I) + K_0(\lambda_Fr_I)]}{I_0(\lambda_Fr_I)[\Theta_{AB}I_0(\lambda_F r _0) + K_0(\lambda_F r_0)]}  \right)^2,
\label{tau2d}
\end{equation}
where $I_0(x)$ and $K_0(x)$ are modified Bessel functions, and
\begin{equation}
\Theta =\frac{\lambda_FI_0(\lambda_Ir_I)K_1(\lambda_Fr_I)+\lambda_II_1(\lambda_Ir_I)K_0(\lambda_Fr_I)}{\lambda_FI_0(\lambda_Ir_I)I_1(\lambda_Fr_I) - \lambda_I I_1(\lambda_Ir_I)I_0(\lambda_Fr_I)}.
\end{equation}

For three dimensions,
\begin{equation}
\frac{T_1}{T_2} = \left(\frac{i_0(\lambda_Fr_0)[\Theta_{AB}i_0(\lambda_Fr_I) + k_0(\lambda_Fr_I)]}{i_0(\lambda_Fr_I)[\Theta_{AB}i_0(\lambda_F r _0) + k_0(\lambda_F r_0)]}  \right)^2,
\label{tau3d}
\end{equation}
where $i_0(x)$ and $k_0(x)$ are spherical modified Bessel functions, and
\begin{equation}
\Theta =   \frac{\lambda_Fi_0(\lambda_Ir_I)k_1(\lambda_Fr_I)+\lambda_I i_1(\lambda_Ir_I)k_0(\lambda_Fr_I)}{\lambda_F i_0(\lambda_Ir_I)i_1(\lambda_Fr_I) - \lambda_I i_1(\lambda_Ir_I)i_0(\lambda_Fr_I)}.
\label{theta3}
\end{equation}

\begin{figure}
\centerline{\includegraphics[width=7.0cm]{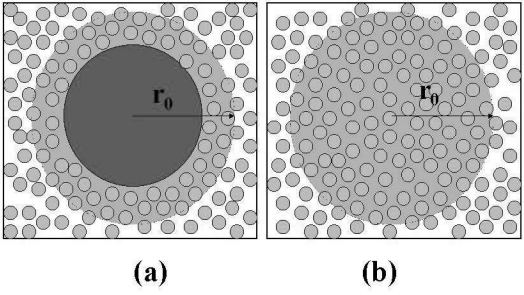}}
\caption{Regions used to calculate the granular temperature. (a) Region around the intruder within a
sphere of radius $r_0$ and (b) region without intruder.}
\label{fig1}
\end{figure}

The granular bed is formed of $N$ monodisperse particles of mass $m_F$ and radius $r_F$. The particles
are inelastic hard disks $(D=2)$ or spheres $(D=3)$. The inelasticity is specified by a restitution
coefficient $e\leq 1$.
The factor $\lambda$ couples the coefficient of thermal conductivity $\kappa = \kappa_0 T_g^{1/2}$ and
the dissipation rate $\gamma = \xi T_g^{3/2}$, explicitly one has\cite{leo}
\begin{equation}
\lambda^2 = \frac{\xi}{2\kappa_0},
\label{lambda}
\end{equation}
where $\xi$ and $\kappa_0$ depend, among other things, on the mass of the particles.
For the fluid particles this is
\begin{equation}
\xi_F = \frac{\Omega_D}{2\sqrt{2\pi}}(1-e^2) n^2 g_0 (2r_F)^{D-1}\left( \frac{2}{m_F} \right)^{1/2},
\label{xi_F}
\end{equation}
and for the region near the intruder
\begin{equation}
\xi_I = \frac{\Omega_D}{2\sqrt{2\pi}}(1-e^2)\frac{n}{V}g_0(r_F+ r_I)^{D-1} \left( \frac{m_I + m_F}{m_I m_F} \right)^{1/2},
\label{xi_I}
\end{equation}
where $n=N/V$, $V=L^D$ is the volume of the system of size $L$, $g_0$ is the pair correlation function for two fluid
particles. In $2D$ the pair correlation function is\cite{Verlet82} 
$g_0 = (1-\frac{7}{16}\nu)/(1-\nu)^2$, with the area fraction $\nu=n\pi r_F^2$.
In $3D$ the pair correlation function is\cite{Carnahan69} $g_0 = (2-\nu)/2(1-\nu)^3$, with
the volume fraction $\nu=4n\pi r_F^3/3$.
Equations (\ref{lambda}), (\ref{xi_F}) and (\ref{xi_I}) define the factors $\lambda_F = \sqrt{\xi_F/2\kappa_0}$ and $\lambda_I = \sqrt{\xi_I/2\kappa_0}$.
The prefactor $\kappa_0$ in two dimensions is
\begin{equation}
\kappa_0 = 3n r_F\left( \frac{\pi}{m_F} \right)^{1/2}\left[ 1 + \frac{1}{3}\frac{1}{G} + \frac{3}{4} \left( 1+\frac{16}{9\pi} \right)G\right],
\end{equation}
and for three dimensions
\begin{equation}
\kappa_0 = \frac{15}{8}n r_F\left( \frac{\pi}{m_F} \right)^{1/2}\left[ 1 + \frac{5}{24}\frac{1}{G}+\frac{6}{5}\left( 1 + \frac{32}{9\pi} \right) G \right],
\end{equation}
where $G=\nu g_0$.

For a dense system the pressure is related to the density by the
virial equation of state
is $p=\frac{1+e}{2}nT_g( 1+\frac{\Omega_D}{2D} n g_0 (2r_F)^D)$.
The thermal expansion coefficient is defined as 
$\alpha\equiv V^{-1} (\partial V /\partial T_g)$.
From the equation of state we can calculate the coefficient  
$\alpha = T_g^{-1}(\nu^2+8)^2/(\nu^3-3\nu^2-8\nu -8)(\nu -1)$ for two dimensions and
$\alpha = T_g^{-1}(\nu^3 -\nu^2-\nu -1)(\nu -1)/(\nu^4-4\nu^3 +4\nu^2+4\nu+1)$
for three dimensions.

The state-dependent viscosity possesses the general form $\mu=\mu_0\sqrt{T_g}$. The
prefactor $\mu_0$ is
\begin{equation}
\mu_0 = \frac{1}{4} n r_F (\pi m_F)^{1/2} \left[ 2 + \frac{1}{G} + \left( 1 + \frac{8}{\pi} \right)G \right],
\end{equation}
for two dimensions and
\begin{equation}
\mu_0 = \frac{1}{3}nr_F(\pi m_F)^{1/2}\left[ 1 + \frac{5}{16}\frac{1}{G} + \frac{4}{5}\left( 1 + \frac{12}{\pi} \right)G \right],
\end{equation}
for three dimensions.

The temperatures of the region with intruder and the region
without intruder are different. 
For $T_1/T_2>1$ the thermal buoyancy force favors the upward movement. When $T_1/T_2<1$ the
thermal buoyancy force favors the downward movement.
The nonlinear form of Equations (\ref{tau2d}) and (\ref{tau3d}) doesn't allow us to calculate analytically
the explicit dependence on the mass ratio $m_I/m_F$ and the size ratio $\phi = r_I/r_F$.  
Let us examine numerically the temperature ratio $T_1/T_2$ as a
function of the mass ratio $m_I/m_F$ for different values of $\phi$. 
We set the number of
particles $N = 5\times 10^3$, the volume fraction $\nu = 0.75$, $r_0=L/2$ and the restitution 
coefficient $e=0.95$.
We consider small particles with unitary mass $m_F = 1$, and we vary the intruder's mass like
$m_I = x m_F$, which $x\sim 0$ $(m_I<<m_F)$ to $x=2$. Figure 2 shows the temperature ratio dependence
as function of the mass ratio for two and three dimensions.
\begin{figure}
\centerline{\includegraphics[width=6.0cm]{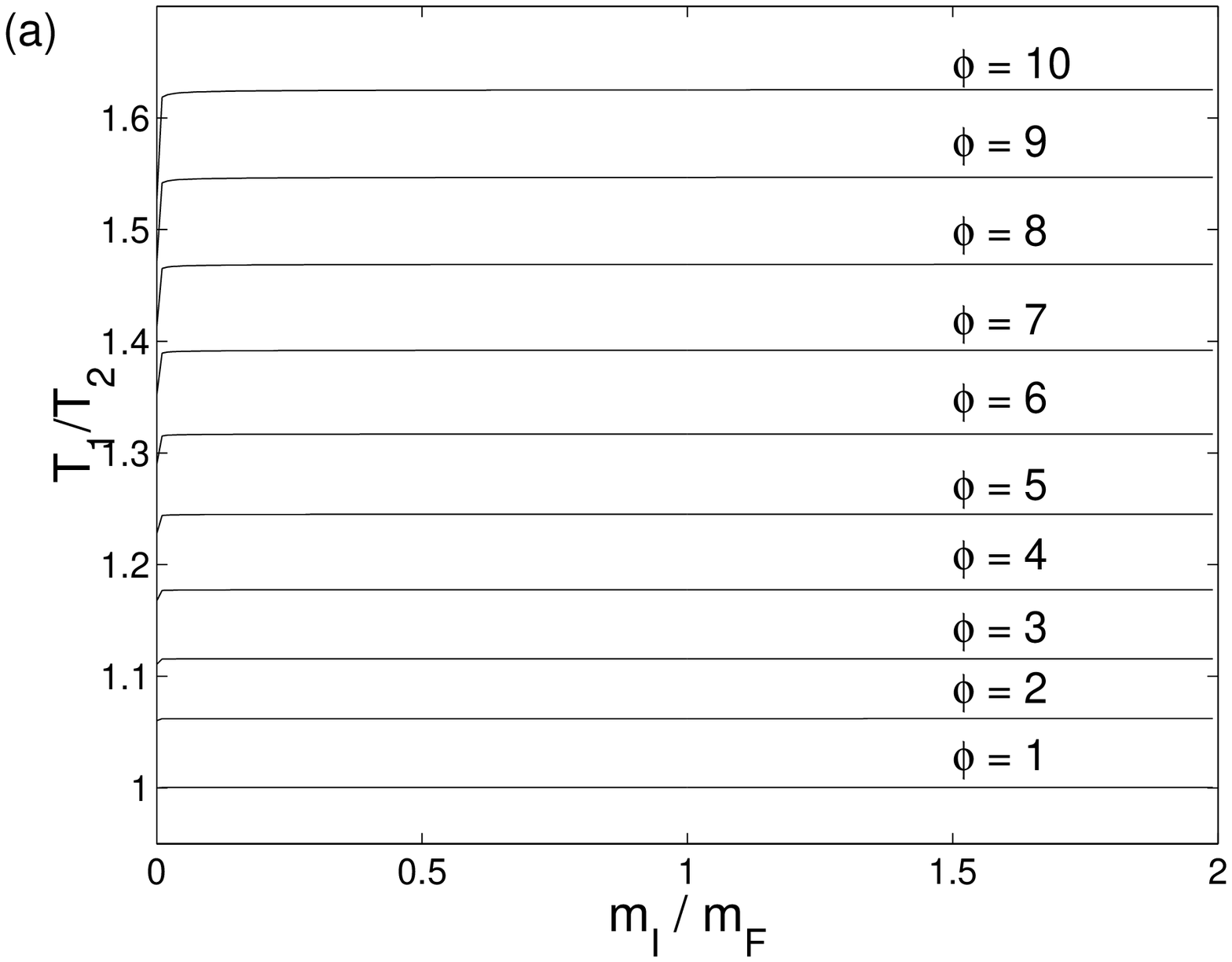}}
\centerline{\includegraphics[width=6.0cm]{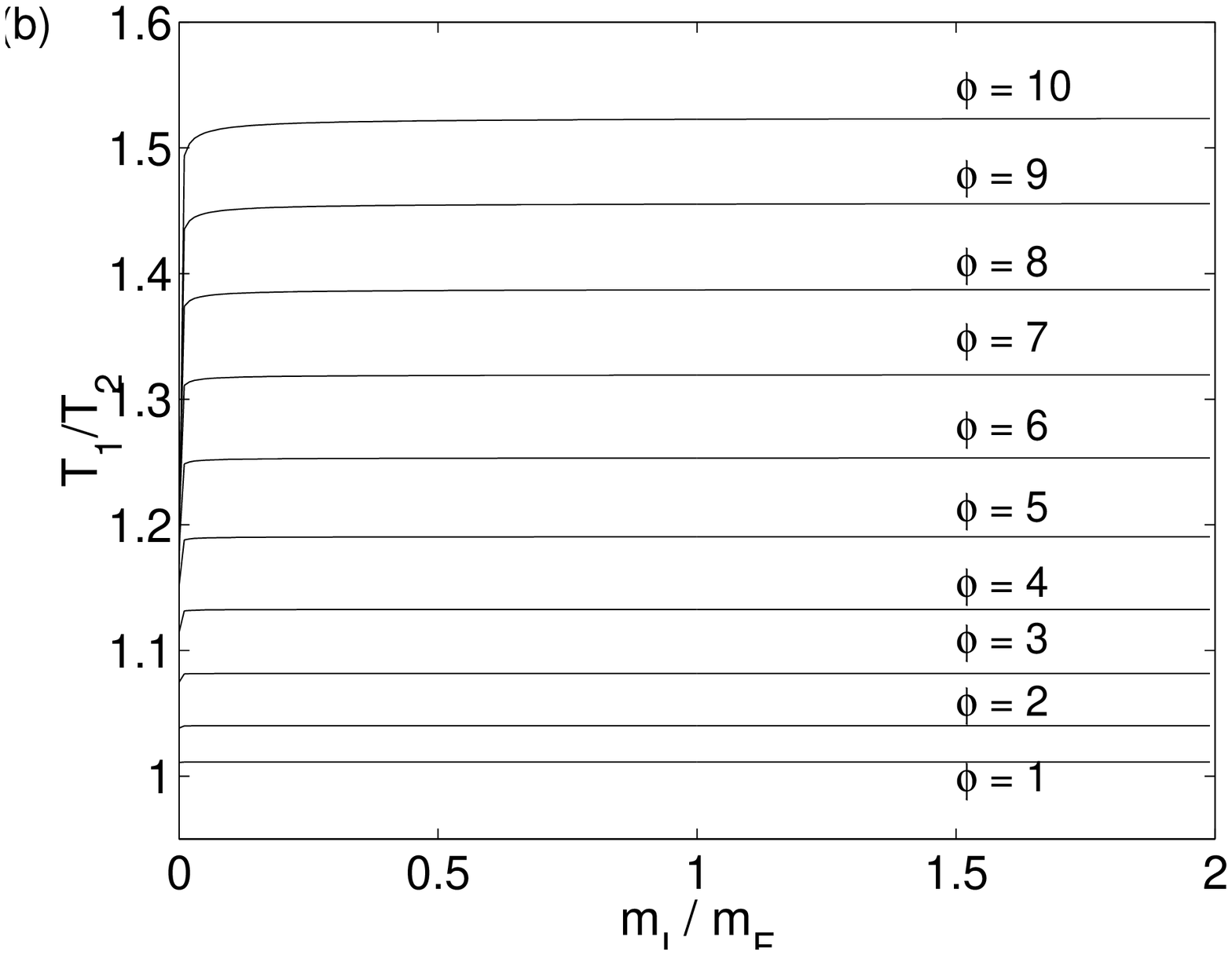}}
\caption{Granular temperature ratio $T_1/T_2$ as a function of the mass ratio $m_I/m_F$ 
for (a) two dimensions, and (b) three dimensions.}
\label{fig2}
\end{figure}

From the analysis of Fig. 2 we can conclude that always $T_1>T_2$. 
and the upward to downward movement is basically controlled  
by the {\it Archimedean buoyancy force}. 

Let us concentrate on the dynamic of the segregation process described by the Equation (\ref{dynamic}). From
the settling velocity $u_s$ (Eq.(\ref{settling})) we can analyze the dependence on size and density ratio. An
uniformly fluidized state can be realized when the granular system is subject to a vertical vibration with
amplitude $A_0$ and frequency $\omega_0=2\pi f$. In the experiments the excitation is described by the
dimensionless acceleration $\Gamma = A_0 \omega_0/g$. The characteristic velocity of the system is 
$u_0 = A_0 \omega_0$. 
The system increases its energy as a result of external driving
while its decreases its energy by dissipation. 
In our theoretical model we do not consider a sinusoidal excitation.
Analogies with shaken granular systems and the dependency of the granular temperature on the amplitude of
vibration has been studied in Ref.\cite{Sunthar99} for dense granular systems in $2D$, in which the
following expression relating the global granular temperature to a symmetric vibration with
maximum velocity $u_0$ is: $T_g = 2\sqrt{2}m_F L (A_0 \omega_0)^2/2Nr_F(1-e^2)$.
For three dimensions we estimate the granular
temperature as $T_g\sim m_F(A_0 \omega_0)^2$. In order to calculate $u_s$ we use the following model 
parameters: mass particle density $\rho_F = 2.7$ gcm$^{-3}$, $r_F = 0.1$ cm, $e = 0.9$, $\nu = 0.75$,
$N = 5\times 10^3$, $g=100$ cms$^{-2}$, $r_0 = L/2$, $A_0 = 2r_F$ cm, $\omega_0 =5.81\sqrt{g/A_0}$
for two dimensions, and  $\omega_0 =0.7\sqrt{g/A_0}$ for three dimensions.
Fig. 3 summarizes the results of our calculations for size ratio $\phi$ from $1$ to $10$ in two dimensions
and three dimensions.

\begin{figure}[h]
\centerline{\includegraphics[width=5.6cm]{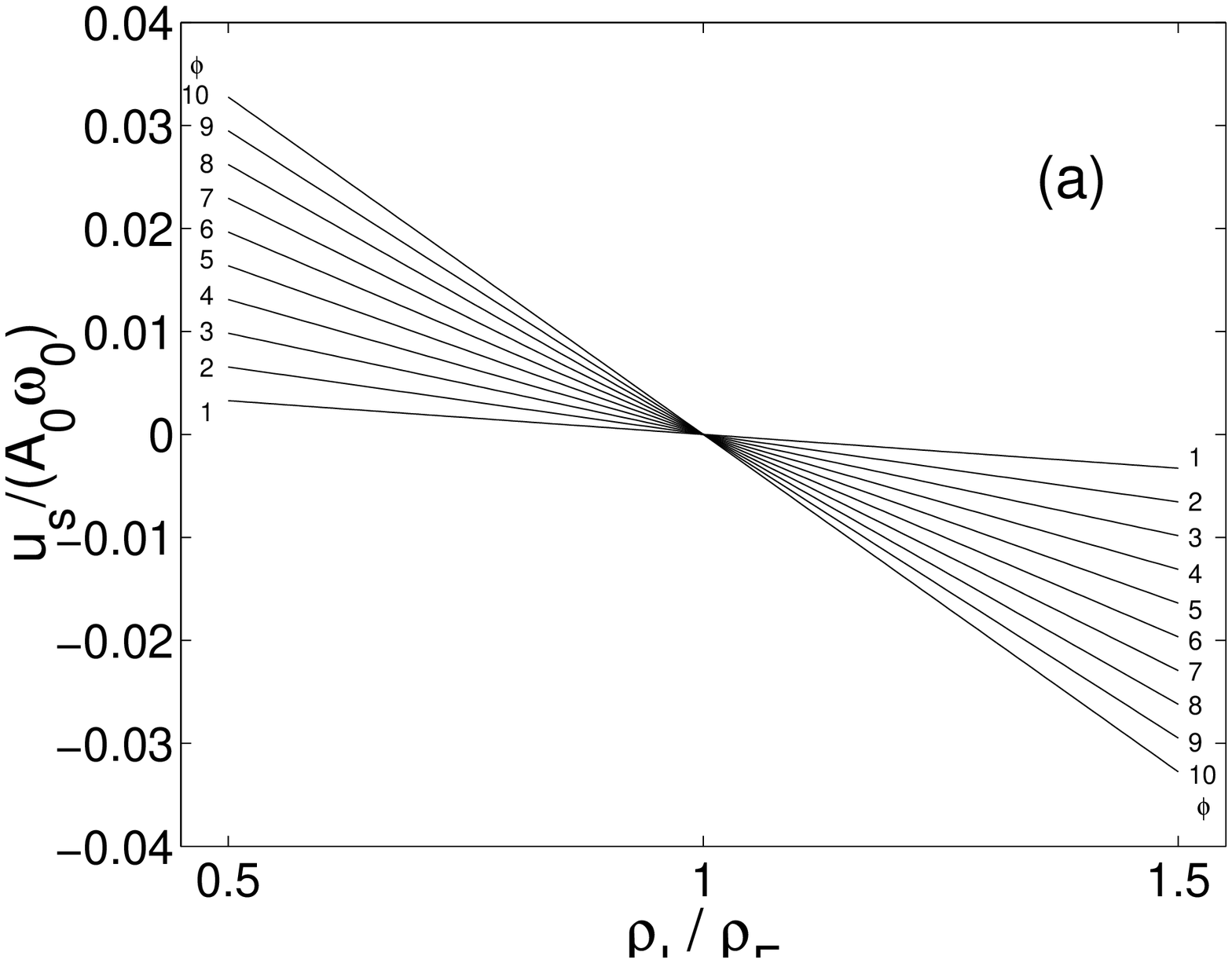}}
\centerline{\includegraphics[width=5.6cm]{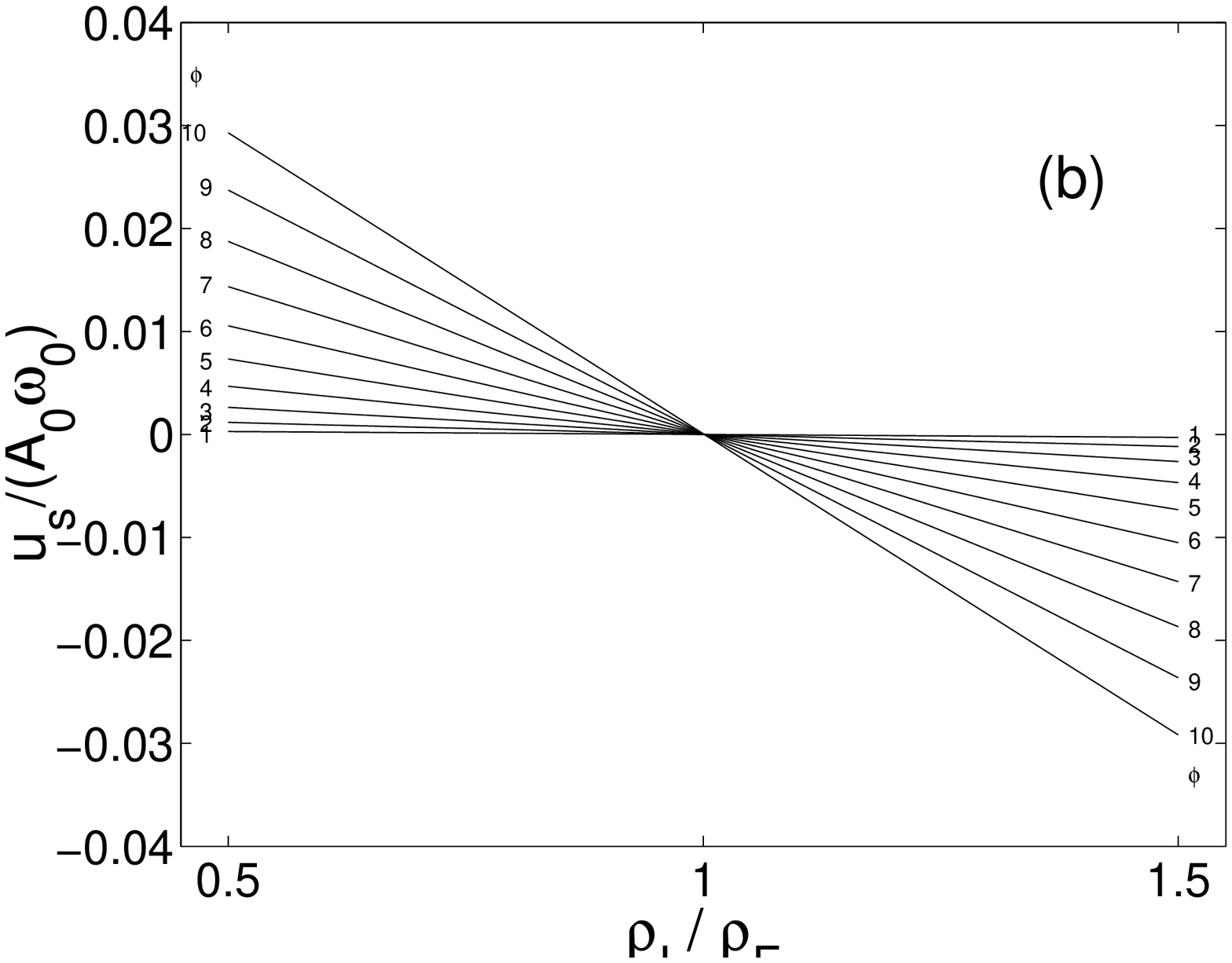}}
\caption{Dimensionless intruder settling velocity $u_s/(A_0\omega_0)$, as a function of density
ratio $\rho_I/\rho_F$ in (a) two dimensions, and (b) three dimensions.}
\label{fig3}
\end{figure}
\begin{figure}[h]
\centerline{\includegraphics[width=5.6cm]{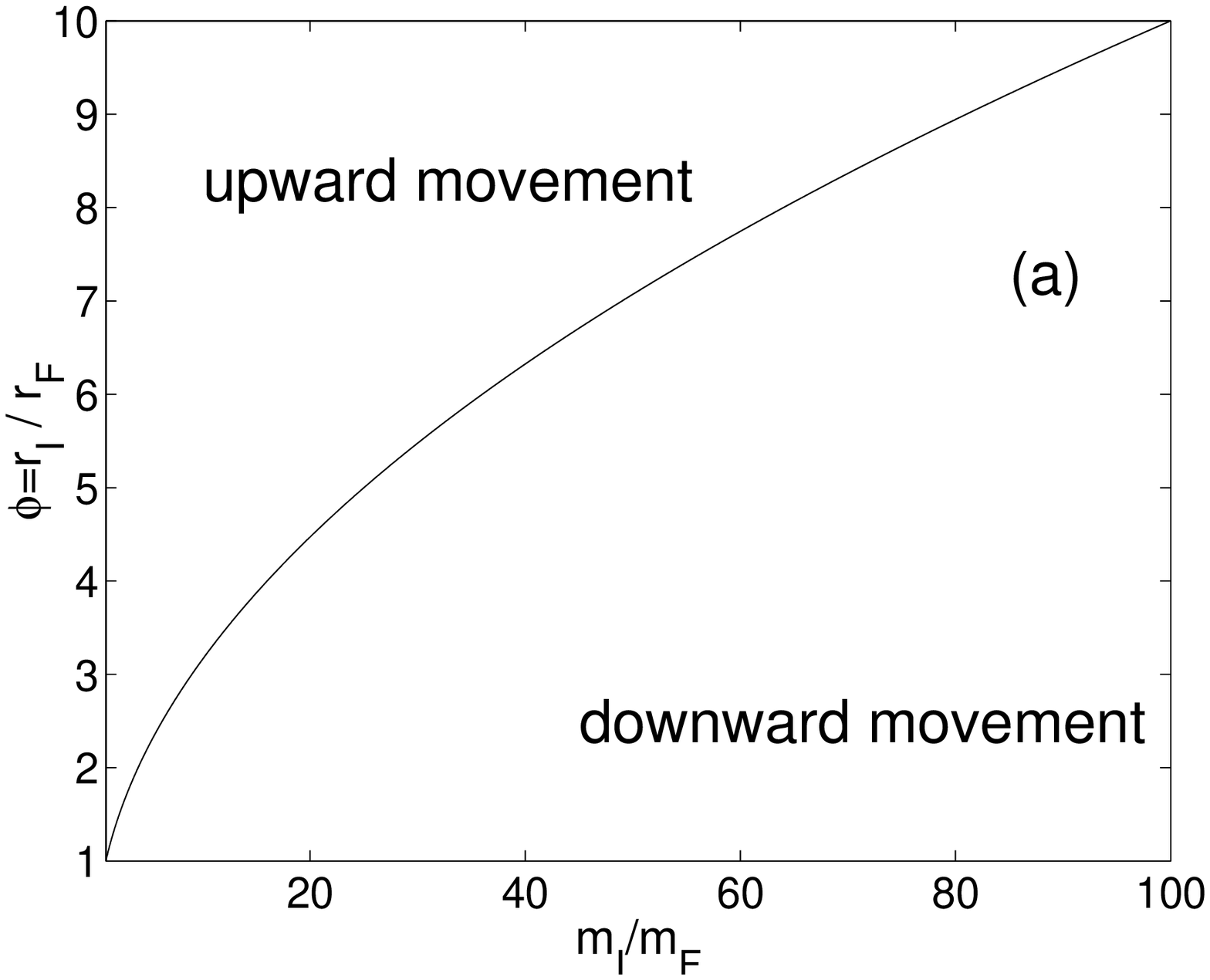}}
\centerline{\includegraphics[width=5.6cm]{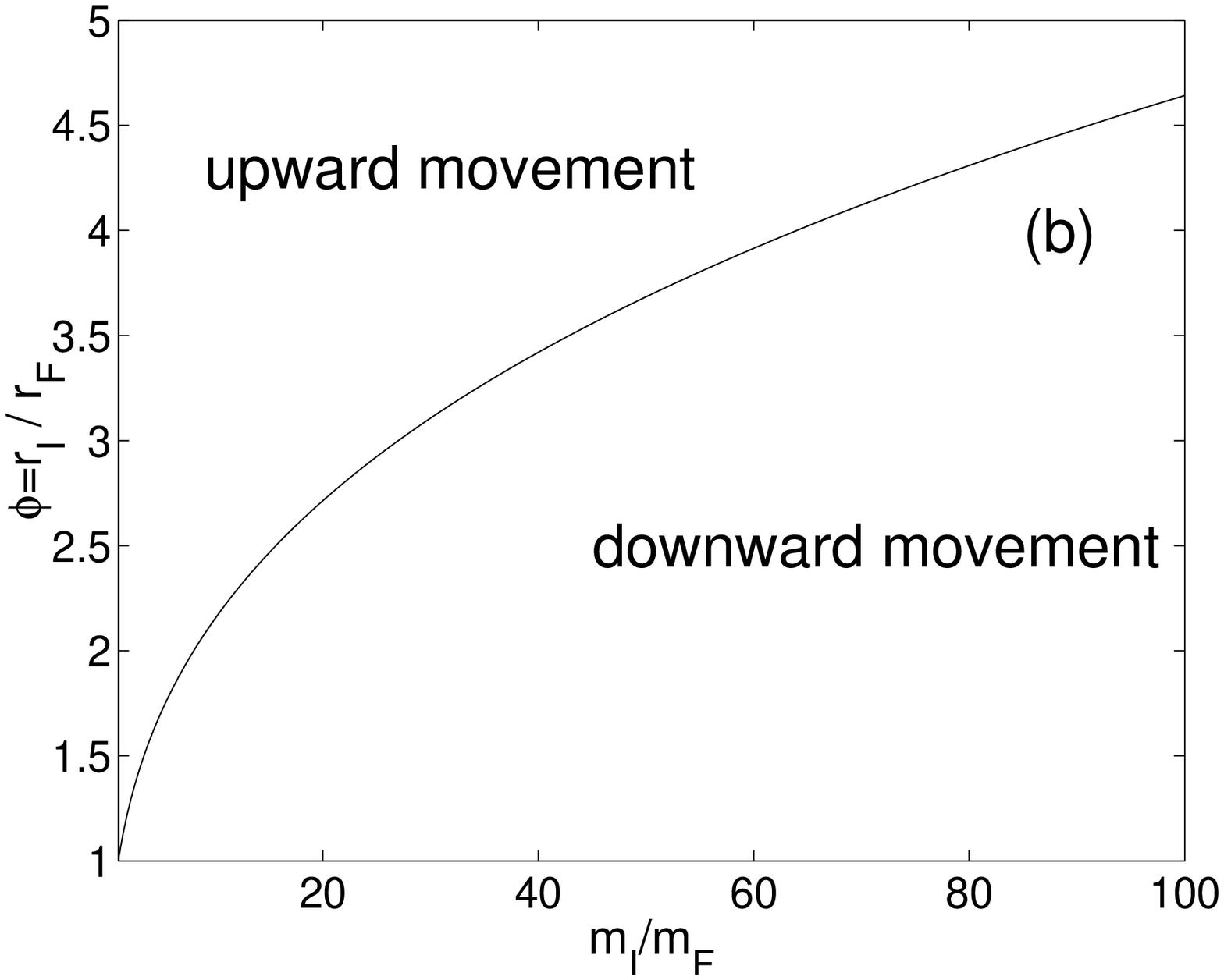}}
\caption{Phase diagram determined for (a) two dimensions, and (b) three dimensions.}
\label{fig4}
\end{figure}

From these results we can observe that the segregation is rapid in systems in which  small particles
are more dense than the intruder. For fixed values of $\phi$, when the density ratio increases the rise
velocity diminishes, this implies a diminution of the intruder rise time. Finally, when the intruder's
density is bigger than the small particles density the intruder particle sinks $u_s<0$. 
For a fixed value of the density ratio the segregation
rate increases with the size ratio. 
These result shown the condition for the crossover from the
upward to the downward intruder's motion. 

Recently, the upward/downward transition has been studied in
references \cite{Hong01,Both02,Jenkins02}. 
In Ref.\cite{Hong01} Hong {\it et al.,} performed molecular dynamic simulations, in two and three
dimensions, of weakly dissipative particles under gravity that are in global thermal
equilibrium with a heat reservoir.
They observed that for a binary mixture of  granular particles the large particles can rise 
and the upward/ downward movement  depends on mass and diameter ratios. 
They proposed an explanation based
on a competition between the percolation effect and the condensation of hard spheres under 
gravity (See Refs.\cite{Hong01,Hong99,Quinn00} for a detailed explanation of the
{\it percolation--condensation} mechanism).
Both and Hong presented a theory based on the variational principle for hard spheres and disks
under gravity\cite{Both02}. They also characterized the segregation phenomenon and investigated the
crossover between the upward/downward movement for large particles and the dependence on mass
and size ratios. In the framework of kinetic theory for a binary granular mixture,
Jenkins and Yoon obtained a segregation criteria for spheres and disks that differ in size
and/or mass\cite{Jenkins02}. Their mechanism is based on a competition between the inertia of the 
particles through the ratio of partial pressures. The three different approaches presented in
references \cite{Hong01,Both02,Jenkins02} coincide qualitatively among them.
In the same spirit we can calculate a phase diagram for the crossover from the upward/downward 
movement (See Figs. 4(a) and 4(b)).
In our model these phase diagrams are derived from the dependence of the segregation velocity on
the mass and size ratio at the situation where the intruder tends to rise ($u_s > 0$), or
to sink ($u_s<0$) (Figs. 3(a) and 3(b)).

In summary, we have studied the segregation dependence on size and density for a single intruder in
a fluidized bed.
On a qualitative level our model is  in agreement with the experimental phenomenology described in
Refs.\cite{Rippie64,Ahmad73}: the higher the density of the intruder the lower the tendency to 
rise. In Fig.(3) we can note that when the density of the particles increases, in the range
where $\rho_I/\rho_F < 1$, the segregation velocity diminishes, so the intruder's rise time will
be bigger. In general a lighter particle rise more quickly than a heavier particle of the same size
in the range where $\rho_I/\rho_F < 1$. In the range where $\rho_I/\rho_F > 1$ the intruder sinks, 
and in this case the downward segregation velocity increases with the density and size.
Our model also agrees with the qualitative behavior reported in Ref.\cite{Shishodia01}, where the 
intruder height decreases as the density ratio increases. In this case, for the range where
$\rho_I/\rho_F < 1$, the height of a lighter particle will be bigger than the height of a
heavier particle of the same size at the same time. This is due to the fact that the rise velocity of
a lighter particle will be bigger than the rise velocity of a heavier particle of the same size.
In agreement with the theoretical results obtained in 
Refs.\cite{Hong01,Both02,Jenkins02} and the numerical simulations of Refs.\cite{Shishodia01,Hong01},
our model predicts that in the case where the density ratio
increases, the intruder particle will sink.

A direct comparison with experiments performed in Ref.\cite{Mobius01} are not appropriate in this work 
since they investigate the intruder's density effect in presence of convection and interstitial air.
Comparison is also difficult with the experiments performed in Refs.\cite{Shinbrot98,Liffman01}, 
where the role of inertia and interstitial air may play an important role. It is important to note
that the experiment performed by Shinbrot and Muzzion\cite{Shinbrot98}, doesn't correspond with
the typical experimental conditions of the Brazil nut effect. In these experiments the intruders
are placed at the surface of a vibrated bed. They reported that the heavy intruder remained at the
surface while the lighter intruder sank. The boundary condition for an intruder in this experimental 
set--up is more different and complicated than the boundary condition for an intruder immersed in
the granular bed. So, in this experiment we can not consider upward/downward 
transition.
In the experiment performed by Liffman {\it et al.,} \cite{Liffman01} they reported that the total rise time
of the intruder is inversely proportional to the density of the intruder. This is, a priori, in contrast to our 
findings. In our case, if the segregation velocity decreases with the intruder's density, then the rise
time decreases. This apparent contradictory fact can be understood from the procedure followed
by Liffman {\it et al.,} for the intruder's density variation.
They studied the motion of disks placed at the base of a vibrated  granular bed. They varied 
the density of the disks with similar size, drilling holes in the disks. The intruder is
``transformed'' from a disk to a ring. The inertia of these two object is different. Also, it is
important to note that the energy dissipation is different. Energy is dissipated during 
collision, among other things, due to the excitation of the internal modes of the object. So,
the coefficient of restitution should be different. In terms of our model, based on the 
concept of granular temperature, the density of dissipated energy should be bigger for a ring than
a disk. In this case the disk is ``hotter'' than a ring, and
the buoyancy force would favor the upward movement.

In order to compare our theory to the experiments one should do measurements varying the intruder's
density homogeneously in dry granular materials. Strangely, a systematic study has not get been reported
in the literature and since this should be easy to carry out, we hope that some experiment and/or
simulation will soon be done.


%
%

%
\end{document}